\documentclass[12pt,a4paper]{article}
\usepackage{graphicx}
\usepackage{amssymb}
\usepackage{amsmath}
\begin{document}
\textwidth=135mm
 \textheight=200mm
\begin{center}
{\bfseries On sensitivity of neutrino-helium ionizing collisions to
neutrino magnetic moments}
\vskip 5mm
K. A. Kouzakov$^{a,}$\footnote{E-mail: kouzakov@srd.sinp.msu.ru} and A. I.
Studenikin$^{a,b,}$\footnote{E-mail: studenik@srd.sinp.msu.ru}
\vskip 5mm {\small {\it $^a$ Faculty of Physics, Lomonosov Moscow State University, 119991 Moscow, Russia}} \\
{\small {\it $^b$ Joint Institute for Nuclear Research, 141980 Dubna,
Russia}}
\\
\end{center}
\vskip 5mm \centerline{\bf Abstract} We consider theoretically ionization
of a helium atom by impact of an electron antineutrino. The sensitivity of
this process to neutrino magnetic moments is analyzed. In contrast to the
recent theoretical prediction, no considerable enhancement of the
electromagnetic contribution with respect to the free-electron case is
found. The stepping approximation is shown to be well applicable
practically down to the ionization threshold.
\vskip 10mm
\section{\label{intro}Introduction}
Electromagnetic properties of neutrinos are of particular interest, for
they open a door to ``new physics'' beyond the Standard Model (SM) (see,
for instance, the review articles~\cite{giunti09,broggini12}). Among these
nontypical neutrino features the most studied and well understood
theoretically are neutrino magnetic moments (NMM). The latter are also
being intensively searched in reactor~\cite{TEXONO,GEMMA},
accelerator~\cite{LSND,DONUT} and solar~\cite{Super-Kamiokande,Borexino}
experiments on low-energy elastic (anti)neutrino-electron scattering. The
current best upper limit on the NMM value obtained in such direct
laboratory measurements is
$$
\mu_\nu\leq2.9\times10^{-11}\mu_B,
$$
where $\mu_B=e/(2m_e)$ is a Bohr magneton. This bound, which is due to the
GEMMA experiment~\cite{GEMMA} with a HPGe detector at Kalinin nuclear
power station, is by an order of magnitude larger than the constraint
obtained in astrophysics~\cite{raffelt90}:
$$
\mu_\nu\leq3\times10^{-12}\mu_B.
$$
And it by many orders of magnitude exceeds the value derived in the
minimally extended SM with right-handed neutrinos~\cite{fujikawa80}
$$
\mu_\nu\leq3\times10^{-19}\mu_B \left(\frac{m_\nu}{1\,{\rm eV}}\right),
$$
where $m_\nu$ is a neutrino mass. At the same time, there are different
theoretical scenarios beyond SM that predict much higher $\mu_\nu$ values,
thus giving hope to observe NMM experimentally in the not too distant
future. Therefore, the major task faced by experiments is to enhance their
sensitivity to the $\mu_\nu$ value.

The strategy of experiments searching for NMM is as follows. One studies
an inclusive cross section for (anti)neutrino-electron scattering which is
differential in the energy transfer $T$. In the ultrarelativistic limit
$m_\nu\to0$, it is given by an incoherent sum of the SM contribution,
which is due to weak interaction that conserves the neutrino helicity, and
the helicity-flipping contribution, which is due to $\mu_\nu$,
\begin{eqnarray}
\label{cr_sec}\frac{d\sigma}{dT}=\frac{d\sigma_{\rm
SM}}{dT}+\frac{d\sigma_{(\mu)}}{dT}.
\end{eqnarray}
In the case of reactor experiments, where one deals with electron
antineutrinos, the SM term is given by
\begin{eqnarray}
\label{cr_sec_SM}\frac{d\sigma_{\rm
SM}}{dT}=\frac{G_F^2m_e}{2\pi}\left[(g_V+g_A)^2+(g_V-g_A)^2\left(1-\frac{T}{E_\nu}\right)^2+(g_A^2-g_V^2)\frac{m_eT}{E_\nu^2}\right],
\end{eqnarray}
where $E_\nu$ is the incident antineutrino energy, $g_A=-1/2$ and
$g_V=(4\sin^2\theta_W+1)/2$, with $\theta_W$ being the Weinberg angle. The
$\mu_\nu$ cross section is given by~\cite{domogatskii70,vogel89}
\begin{eqnarray}
\label{cr_sec_mu}\frac{d\sigma_{(\mu)}}{dT}=4\pi\alpha\mu_\nu^2\left(\frac{1}{T}-\frac{1}{E_\nu}\right),
\end{eqnarray}
where $\alpha$ is the fine-structure constant. Thus, the two components of
the cross section~(\ref{cr_sec}) exhibit quite different dependencies on
the recoil-electron kinetic energy $T$. Namely, at low $T$ values the SM
cross section is practically constant in $T$, while that due to $\mu_\nu$
behaves as $1/T$. This means that the experimental sensitivity to NMM
value critically depends on lowering the energy threshold of the detector
employed for measurement of the recoil-electron spectrum.

The formulas~(\ref{cr_sec_SM}) and~(\ref{cr_sec_mu}) assume the electron
to be free and initially at rest. The energy threshold reached so far in
the aforementioned GEMMA experiment with a HPGe detector is
2.8\,keV~\cite{GEMMA}. This value is already much lower than the binding
energy of $K$-electrons in Ge atoms ($\sim10$\,keV). This fact makes it
necessary to take into account the atomic effects beyond the free-electron
(FE) approximation. The results of the corresponding treatment performed
in~\cite{wong10} suggested that the electron binding in atoms can
dramatically increase the $\mu_\nu$ contribution to the differential cross
section~(\ref{cr_sec}) as compared with the FE case. However, the careful
and detailed theoretical analysis~\cite{plb11,jpl11,prd11} has found no
evidence of the claimed ``atomic ionization effect''. Moreover, it
provided general arguments supporting the so-called stepping approximation
formulated in~\cite{kopeikin97} on the basis of numerical calculations for
various targets. According to the stepping approximation, the cross
section $d\sigma/dT$ for knocking-out an electron from an atomic orbital
follows the FE dependence on $T$ all the way down to the ionization
threshold $T_I$ for this orbital with a very small (at most a few percent)
deviation. And the orbital becomes ``inactive'' when $T<T_I$, thus
producing a sharp step in the $T$ dependence of $d\sigma/dT$ summed over
all occupied atomic levels.

Recently, the authors of~\cite{martemyanov11} deduced by means of
numerical calculations that the $\mu_\nu$ contribution to ionization of
the He target by impact of electron antineutrinos from reactor and tritium
sources strongly departures from the stepping approximation, exhibiting
large enhancement relative to the FE approximation. According
to~\cite{martemyanov11}, the effect is maximal when the $T$ value
approaches the ionization threshold in helium, $T_I=24.5874$\,eV, where
the relative enhancement is as large as almost eight orders of magnitude.
It was thus suggested that this finding might have an impact on searches
for $\mu_\nu$, provided that its value falls within the range
$10^{-13}-10^{-12}\mu_B$. The purpose of the present Letter is to show
that (i) the result of~\cite{martemyanov11} is erroneous and (ii) the
stepping approximation for helium is well applicable, except the energy
region $T\sim T_I$ where the differential cross section substantially
decreases relative to the FE case.
\section{\label{theo}Theory of neutrino-impact ionization of helium}
We consider the process where an electron antineutrino with energy $E_\nu$
scatters on a He atom at energy and spatial-momentum transfers $T$ and
${\bf q}$, respectively. In what follows we focus on the ionization
channel of this process in the kinematical regime $T\ll E_\nu$, which
mimics a typical situation with reactor ($E_\nu\sim1$\,MeV) and tritium
($E_\nu\sim10$\,keV) antineutrinos when the case $T\to T_I$ is concerned.
The He target is assumed to be in its ground state $|\Phi_i\rangle$ with
the corresponding energy $E_i$. Since for helium one has $\alpha Z\ll1$,
where $Z=2$ is the nuclear charge, the state $|\Phi_i\rangle$ can be
treated nonrelativistically. As we are interested in the energy region
$T\sim T_I$, the final He state $|\Phi_f\rangle$ (with one electron in
continuum) can also be treated in the nonrelativistic approximation.

Under the above assumptions, the SM and $\mu_\nu$ components of the
differential cross section for the discussed ionization process can be
presented as~\cite{prd11}
\begin{eqnarray}
\label{cr_sec_He_SM} \frac{d\sigma_{\rm
SM}}{dT}&=&\frac{G_F^2}{4\pi}(1+4\sin^2{\theta_W}+8\sin^4{\theta_W})\int_{T^2}^{4
E_\nu^2}S(T,q^2)\,dq^2,\\
\label{cr_sec_He_mu} \frac{d\sigma_{(\mu)}}{dT}&=&4 \pi \alpha \mu_\nu^2
\int_{T^2}^{4 E_\nu^2} S(T,q^2)\,\frac{dq^2}{q^2},
\end{eqnarray}
where $S(T,q^2)$ is the dynamical structure factor given by
\begin{eqnarray}
\label{dyn_str_fact} S(T,q^2)=\sum_f\left|\langle
\Phi_f(\textbf{r}_1,\textbf{r}_2) |e^{i{\bf qr}_1} + e^{i{\bf
qr}_2}|\Phi_i(\textbf{r}_1,\textbf{r}_2)\rangle\right|^2
\delta(T-E_f+E_i).
\end{eqnarray}
Here the $f$ sum runs over all final He states having one electron ejected
in continuum, with $E_f$ being their energies.

For evaluation of the dynamical structure factor~(\ref{dyn_str_fact}) we
employ the same models of the initial and final He states as
in~\cite{martemyanov11}. The initial state is given by a product of two
$1s$ hydrogenlike wave functions with an effective charge $Z_{i}$,
\begin{eqnarray}
\label{init_state}\Phi_i(\textbf{r}_1,\textbf{r}_2)=\varphi_{1s}(Z_{i},\textbf{r}_1)\varphi_{1s}(Z_{i},\textbf{r}_2),
\qquad \varphi_{1s}(Z_{i},\textbf{r})=\sqrt{\frac{Z_{i}^3}{\pi
a_0^3}}\,e^{-Z_{i}r/{a_0}},
\end{eqnarray}
where $a_0=1/(\alpha m_e)$ is the Bohr radius. The final state has the
form
\begin{equation}
\label{fin_state} \Phi_f(\mathbf{r}_1,\mathbf{r}_2)=\frac{1}{\sqrt{2}}[
\varphi_{\mathbf{k}}^-(Z_f,\mathbf{r}_1)\varphi_{1s}(Z,\mathbf{r}_2)+\varphi_{\mathbf{k}}^-(Z_f,\mathbf{r}_2)\varphi_{1s}(Z,\mathbf{r}_1)],
\end{equation}
where $\varphi_{\mathbf{k}}^-(Z_f,\mathbf{r})$ is an outgoing Coulomb wave
for the ejected electron with spatial momentum ${\bf k}$. $Z_f$ is the
effective charge experienced by the ejected electron in the field of the
final He$^+$ ion. Contributions to the dynamical structure factor from
excited He$^+$ states are neglected due to their very small overlap with
the $K$-electron state in the He atom.

To avoid nonphysical effects connected with nonorthogonality of
states~(\ref{init_state}) and~(\ref{fin_state}), we use the Gram-Schmidt
orthogonalization
$$
|\Phi_f\rangle\rightarrow|\Phi_f\rangle-\langle\Phi_i|\Phi_f\rangle|\Phi_i\rangle.
$$
Substitution of~(\ref{init_state}) and~(\ref{fin_state})
into~(\ref{dyn_str_fact}) thus yields
\begin{eqnarray}
\label{dyn_str_fact_1}S(T,q^2)=\int\frac{d{\bf k}}{(2\pi)^3} |F({\bf
k},{\bf q})|^2
\delta\left(T-\frac{k^2}{2m_e}+2\alpha^2m_e-Z_{i}^2\alpha^2m_e\right),
\end{eqnarray}
where $k=\sqrt{2m_e(T+2\alpha^2m_e-Z_{i}^2\alpha^2m_e)}$, and
\begin{equation}
F({\bf k},{\bf q})=\sqrt{2}\langle \varphi_{\bf
k}^-(Z_f,\textbf{r}_1)\varphi_{1s}(Z,\textbf{r}_2)| e^{i{\bf qr}_1} +
e^{i{\bf qr}_2}-2\rho_{1s}({\bf q})|\varphi_{1s}(Z_{i},{\bf
r}_1)\varphi_{1s}(Z_{i},{\bf r}_2)\rangle
\end{equation}
is the inelastic form factor, with
\begin{equation}
\rho_{1s}({\bf q})=\int\varphi_{1s}(Z_{i},{\bf r})e^{i{\bf
qr}}\varphi_{1s}(Z_{i},{\bf r})\,d{\bf r}.
\end{equation}
It is straightforward to perform the further calculation of the dynamical
structure factor analytically\footnote{The resulting expressions are
omitted for the sake of brevity.} (see, for instance, the
textbook~\cite{landafshiz}).

Finally, the usual choice of the effective charges is
$Z_i=27/16\approx1.69$ and $Z_f=1$ (see, for instance,~\cite{pra12} and
references therein). The value $Z_i=27/16$ follows from the variational
procedure that minimizes the ground-state energy $E_i$, while the value
$Z_f=1$ ensures the correct asymptotic behavior of the final state.
However, the authors of~\cite{martemyanov11} utilized in their
calculations the values $Z_i=1.79$ and $Z_f=1.1$ derived from fitting the
photoionization cross-section data on helium with the present model of the
He states.
\section{\label{res}Results and discussion}
The departures of the differential cross sections~(\ref{cr_sec_He_SM})
and~(\ref{cr_sec_He_mu}) from the FE approximation are characterized by
the respective atomic factors
\begin{eqnarray}
\label{at_factor}f_{\rm SM}=\frac{d\sigma_{\rm SM}/dT}{d\sigma_{\rm
SM}^{\rm FE}/dT}, \qquad f_{\rm
NMM}=\frac{d\sigma_{(\mu)}/dT}{d\sigma_{(\mu)}^{\rm FE}/dT},
\end{eqnarray}
where $d\sigma_{\rm SM}^{\rm FE}/dT$ and $d\sigma_{(\mu)}^{\rm FE}/dT$ are
the SM and $\mu_\nu$ contributions to the differential cross section for
scattering of an electron antineutrino on two free electrons. Let us
recall that following~\cite{martemyanov11} one should expect the $f_{\rm
NMM}$ value to be of about $10^8$ at $T\to T_I$.
\begin{figure}
\begin{center}
\includegraphics[width=0.9\textwidth]{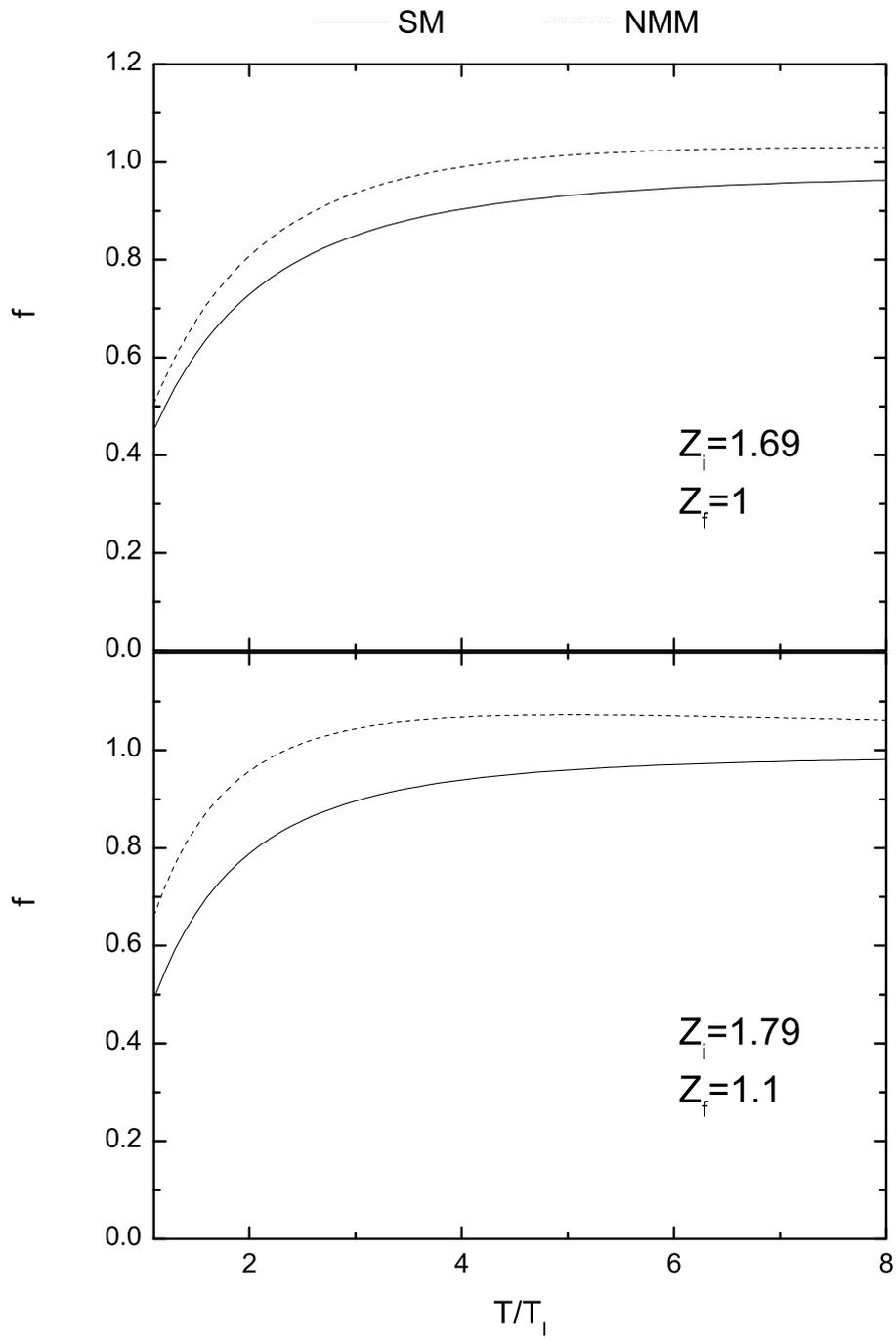}
\end{center}
\caption{\label{fig}Atomic factors as functions of the energy transfer.}
\end{figure}

Numerical results for atomic factors~(\ref{at_factor}) are shown in
Fig.~\ref{fig}. They correspond to the kinematical regime $T\ll\alpha
m_e\ll 2E_\nu$, which is typically realized both for reactor and for
tritium antineutrinos when $T<200$\,eV. Note that in such a case one can
safely set the upper limit of integrals in~(\ref{cr_sec_He_SM})
and~(\ref{cr_sec_He_mu}) to infinity, as the dynamical structure factor
$S(T,q^2)$ rapidly falls down when $q\gtrsim\alpha m_e$ and practically
vanishes in the region $q\gg\alpha m_e$. It can be seen from
Fig.~\ref{fig} that atomic factors exhibit similar behaviors for both sets
of the $Z_i$ and $Z_f$ parameters discussed in the previous section.
Namely, their values are minimal ($\sim0.5$) at the ionization threshold,
$T=T_I$, and tend to unity with increasing $T$. The latter tendency is
readily explained by approaching the FE limit. It can be also seen that a
more or less serious deviation ($>10\%$) of the present results from the
stepping approximation is observed only in the low-energy region
$T<100$\,eV.

Thus, the present calculations do not confirm the huge enhancement of the
$\mu_\nu$ contribution with respect to the FE approximation. Moreover, in
accord with various calculations for other atomic
targets~\cite{plb11,jpl11,prd11,kopeikin97,dobretsov92,fayans92,fayans01,kopeikin03},
we find that at small energy-transfer values the electron binding in
helium leads to the appreciable reduction of the differential cross
section relative to the FE case. We attribute the erroneous prediction
of~\cite{martemyanov11} to the incorrect dynamical model that draws an
analogy between the NMM-induced ionization and photoionization. Indeed, as
discussed in~\cite{plb11}, the virtual photon in the NMM-induced
ionization process can be treated as real only when $q\to T$. However, the
integration in~(\ref{cr_sec_He_mu}) involves the $q$ values ranging from
$T$ up to $2E_\nu$. Since $E_\nu\gg T$, the real-photon picture appears to
be applicable only in the vicinity of the lower integration limit. When
moving away from that momentum region, one encounters a strong departure
from the real-photon approximation which treats the integrand as a
constant in the whole integration range, assuming it to be equal to its
value at $q=T$, that is,
$$
\frac{1}{q^2}\,S(T,q^2)=\frac{1}{T^2}\,S(T,T^2).
$$
Such an approach is manifestly unjustified, and it gives rise to the
spurious enhancement of the $\mu_\nu$ contribution to the differential
cross section.
\section{\label{sum}Summary}
We carried out a theoretical analysis of ionization of helium by
electron-antineutrino impact. Our calculations showed no evidence of the
enhancement of the electromagnetic contribution as compared with the FE
case. In contrast, in line with previous studies on other targets, we
found that the magnitudes of the differential cross sections decrease
relative to the FE approximation when the energy transfer is close to the
ionization threshold. Thus, no sensitivity enhancement can be expected
when using the He target in searches for NMM. And the stepping
approximation appears to be valid, within a few-percent accuracy, down to
the energy-transfer values as low as almost 100\,eV.

{\bf Acknowledgements.} We are grateful to A. S. Starostin for useful
discussions.
\end{document}